\documentclass[a4paper,11pt]{article}
\pdfoutput=1 
\usepackage{jheppub} 
\usepackage[T1]{fontenc} 
\usepackage{amsmath,amssymb,epsfig,txfonts,relsize}
\usepackage{xspace}

\def\etmiss{\ensuremath{E_{T}^{\mathrm{miss}}}\xspace}

\def\ptmiss{\ensuremath{\vec p^{\mathrm{\ miss}}_T\xspace}}

\def\TeV{\ifmmode {\mathrm{\ Te\kern -0.1em V}}\else
                   \textrm{Te\kern -0.1em V}\fi}
\def\GeV{\ifmmode {\mathrm{\ Ge\kern -0.1em V}}\else
                   \textrm{Ge\kern -0.1em V}\fi}

\def \beq{\begin{equation}}
\def \eeq{\end{equation}}
\def \bea{\begin{eqnarray}}
\def \eea{\end{eqnarray}}

\begin{document}

\title{Searching for dark matter  in  final states  \\ with two jets and missing  transverse energy}

\author[1]{Ulrich Haisch}

\author[2]{and Giacomo Polesello}

\affiliation[1]{Max Planck Institute for Physics, F{\"o}hringer Ring 6,  80805 M{\"u}nchen, Germany}

\affiliation[2]{INFN, Sezione di Pavia, Via Bassi 6, 27100 Pavia, Italy}

\emailAdd{haisch@mpp.mpg.de}
\emailAdd{giacomo.polesello@cern.ch}

\abstract{We reemphasise the usefulness of angular correlations in LHC searches for missing  transverse energy~($E_T^{\mathrm{miss}}$) signatures that involve jet  $(j)$ pairs with large invariant mass. For the case of mono-jet production via  gluon-fusion, we develop a realistic analysis strategy that allows to split the dark matter~(DM) signal into distinct one jet-like and two jet-like event samples. By performing state-of-the-art Monte Carlo simulations of both the mono-jet signature and the standard model background, it is shown that the dijet  azimuthal angle difference $\Delta \phi_{j_1 j_2}$  in $2 j + \etmiss$ production provides a powerful discriminant in realistic  searches. Employing  a shape fit  to  $\Delta \phi_{j_1 j_2}$, we  then determine the  LHC reach of the mono-jet channel in the context of spin-0 $s$-channel DM simplified models. The constraints obtained by the proposed  $\Delta \phi_{j_1 j_2}$ shape fit turn out to be significantly more stringent than those that derive from standard $\etmiss$ shape analyses.}

\maketitle
\flushbottom

\section{Introduction}

One of the main channels used in the search for direct production of dark matter~(DM) at hadron colliders is the final state that includes a high transverse momentum~($p_T$)  hardonic jet recoiling against the undetectable DM particles. These so-called mono-jet searches have a long and partly colourful history.  They have been performed in the past at  all  major general purpose detector experiments like  UA1, CDF and D{\O}~\cite{Arnison:1984qu,Abazov:2003gp,Aaltonen:2008hh}, and at the LHC they now represent an important  pillar of the search strategy for new physics beyond the standard model~(SM). 

The latest ATLAS and CMS mono-jet results~\cite{Aaboud:2017phn,Sirunyan:2017jix} are based on an integrated luminosity of around $36 \, {\rm fb}^{-1}$ of $13 \, {\rm TeV}$ LHC data.  Both searches require the presence of at least a single  high-$p_T$ jet and fit the shape of the missing transverse energy~(\etmiss) spectrum to extract limits on DM production. Given the good theoretical understanding of the $Z/W+{\rm jets}$ SM backgrounds~\cite{Lindert:2017olm}, LHC~measurements of the $\etmiss$ distribution in mono-jet production place the leading constraints on $s$-channel DM simplified models~(see~for instance~\cite{Abdallah:2015ter,Abercrombie:2015wmb,Boveia:2016mrp} and references therein) in certain regions of parameter space. However, since the corresponding shapes of the $\etmiss$  spectra are  featureless and largely independent of the type of  mediation mechanism, existing mono-jet measurements provide  insufficient information to determine additional DM properties. 

In order to  disentangle different types of DM-SM interactions through  studies of  $X +\etmiss$ signatures more complicated observables and/or final states~$X$ need to be considered. The simplest option are channels with at least  two SM particles such  as two jets~($j$) or two charged~leptons~($l$) and $\etmiss$ in the final state. The~$2 j + \etmiss$ signature can thereby result from either the gluon-fusion~\cite{Haisch:2013fla} or the vector-boson fusion~\cite{Eboli:2000ze,Cotta:2012nj,Crivellin:2015wva}  channel, while the $2 l+ \etmiss$ signal can for instance arise  from  $t \bar t + \etmiss$~\cite{Buckley:2015ctj,Haisch:2016gry} and $t W + \etmiss$~\cite{181111048,Haisch:2018tw} production. In~all cases the angular correlations of the visible final state particles have been studied and shown to provide  useful information on the structure of the interactions between the dark and the SM sector. The goal of this work is to reassess the usefulness of dijet angular correlations in the gluon-fusion $2 j + \etmiss$ channel, applying the general ideas presented in~\cite{Haisch:2013fla} to the case of spin-0 simplified DM models. By~performing simulations of the DM signal taking into account  the effects of matrix element matching and merging, parton shower and hadronisation corrections and a realistic detector modelling, we show that the azimuthal angle difference~$\Delta \phi_{j_1 j_2}$ between the two jets in $2 j + \etmiss$ events furnishes a powerful model discriminant in a realistic  experimental environment. We find that compared to standard~$\etmiss$  likelihood fits,  the inclusion of shape information on $\Delta \phi_{j_1 j_2}$  should lead to a significantly improved reach  in  spin-0 $s$-channel DM simplified models. Projections are presented based on $300 \,{\rm fb}^{-1}$ and $3 \, {\rm ab}^{-1}$ of 14~TeV LHC data, corresponding to LHC Run-3 and the high-luminosity phase of the LHC~(HL-LHC).

Our article is organised as follows. In Section~\ref{sec:generation} we briefly describe  the structure of the~DM simplified models that we use to interpret the $2j+\etmiss$ searches. In this section also the generation of the DM signal and the SM backgrounds is explained and our detector simulation is described.  Section~\ref{sec:analysis} details our analysis strategy and discusses  the angular correlations of the DM signal. The LHC Run-3 and  HL-LHC projections are presented in Section~\ref{sec:results}. We conclude in Section~\ref{sec:conclusions}. Supplementary material can be found in Appendix~\ref{sec:appendix}. 

\section{DM signal and SM backgound}
\label{sec:generation}

In our work the following simplified Lagrangians are studied~(see e.g.~\cite{Abdallah:2015ter,Abercrombie:2015wmb,Boveia:2016mrp})
\beq \label{eq:lagrangians}
\begin{split}
{\cal L}_\phi & \supset -g_\chi \hspace{0.25mm} \phi  \hspace{0.25mm}  \bar \chi \chi - \frac{\phi}{\sqrt{2}} \sum_{q=u,d,s,c,b,t}  g_q  \hspace{0.25mm}  y_q  \hspace{0.25mm} \bar q q \,, \\[2mm]
{\cal L}_a & \supset -i g_\chi  \hspace{0.25mm} a  \hspace{0.25mm}  \bar \chi  \hspace{0.25mm} \gamma_5  \hspace{0.25mm} \chi - i  \hspace{0.25mm} \frac{a}{\sqrt{2}} \sum_{q=u,d,s,c,b,t}  g_q  \hspace{0.25mm}  y_q  \hspace{0.25mm} \bar q  \hspace{0.25mm} \gamma_5  \hspace{0.25mm} q \,,
\end{split}
\eeq
which describe the coupling of a dark sector to the SM through the $s$-channel exchange of scalar~($\phi$) and pseudoscalar~($a$) mediators. In~(\ref{eq:lagrangians}) the symbol $\chi$ represents the DM particle assumed to be a Dirac fermion, $g_\chi$ is a dark-sector Yukawa coupling,  $y_q = \sqrt{2} m_q/v$ are the SM quark Yukawa couplings with~$m_q$ the mass of the relevant quark $q$ and $v \simeq 246 \, {\rm GeV}$ the Higgs vacuum expectation value, and $\gamma_5$ denotes the fifth Dirac matrix. 

\begin{figure}[t!]
\begin{center}
\includegraphics[width=0.75\textwidth]{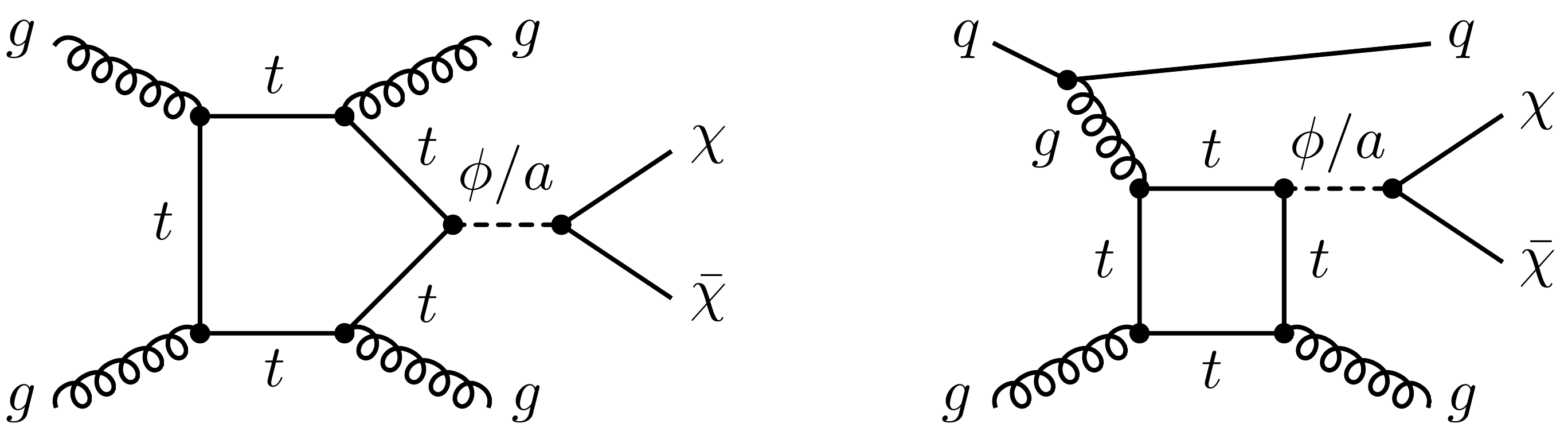}
\vspace{2mm}
\caption{Two examples of Feynman diagrams describing $2j+\etmiss$ production in the spin-0 $s$-channel DM simplified models~(\ref{eq:lagrangians}). }
\label{fig:diagrams}
\end{center}
\end{figure}

The DM signal samples are generated at leading order~(LO) using  the {\tt DMsimp}~\cite{Backovic:2015soa}   implementation of the Lagrangians~(\ref{eq:lagrangians}) together with {\tt  MadGraph5\_aMC@NLO}~\cite{Alwall:2014hca} and {\tt NNPDF3.0}~\cite{Ball:2012cx} parton distribution functions~(PDFs). The associated production of DM with one and two jets are generated for $pp$ collisions at a centre-of-mass~(CM) energy of 14~TeV. See~Figure~\ref{fig:diagrams} for representative  one-loop graphs  that contribute to the $2 j +\etmiss$ signal in the  spin-0 $s$-channel DM simplified models. The events are showered with {\tt PYTHIA~8.2}~\cite{Sjostrand:2014zea} using the Catani-Kraus-Kuhn-Webber~(CKKW) jet matching prescription~\cite{Catani:2001cc}. We consider five different values of the mediator mass $M_{\phi/a}$ in the range from $50 \, {\rm GeV}$ to $1 \, {\rm TeV}$.  The mass of the DM particles is set to $m_\chi = 1 \, {\rm GeV}$ and we employ $g_\chi = g_t =1$ for the couplings of the mediators to DM and top quarks. The total decay width $\Gamma_{{\phi}/a}$ of the mediator is assumed to be minimal and calculated at tree level using {\tt  MadGraph5\_aMC@NLO}. Since  in the narrow width approximation  the signal predictions factorise  into the cross sections for $pp \to 2j + \phi/a$ production times the $\phi/a \to \chi \bar \chi$ branching ratio, changing  $\Gamma_{{\phi}/a}$ leads only  to a rescaling of the signal strength. The experimental acceptance is instead insensitive to the total decay width, and hence  it is sufficient to generate samples for a single  choice of couplings.  The predictions for other values of $g_\chi$ and $g_t$ can then be obtained by scaling with the associated $\phi/a \to \chi \bar \chi$ branching ratio.  

The dominant SM backgrounds arise from $Z/W+{\rm jets}$ production. We consider separately the $Z+\mathrm{jets}$ channel with $Z \to \nu \bar \nu$ and the $W+\mathrm{jets}$ mode with $W \to l \nu$ where $l=e,\mu,\tau$. These backgrounds are  generated at LO with {\tt  MadGraph5\_aMC@NLO} and {\tt NNPDF3.0} PDFs, and can contain up to two additional jets  in the matrix element. The generation is performed in slices of the vector-boson $p_T$, and the resulting events are showered with {\tt PYTHIA~8.2} employing   a CKKW jet matching. The inclusive signal region~IM3  of the analysis~\cite{Aaboud:2017phn} requires $\etmiss > 350 \, {\rm GeV}$, and  for these selections the background from $Z/W+{\rm jets}$ production amounts to around 95\% of the total SM background.  Our $Z/W+{\rm jets}$  samples are normalised such that the different contributions match the number of events in the~IM3 signal region as estimated by the ATLAS collaboration scaled to a CM energy of 14~TeV and to the appropriate integrated luminosity. In addition, the small SM backgrounds arising from~$t\bar{t}$~\cite{Campbell:2014kua}, $tW$~\cite{Re:2010bp} and diboson~\cite{Melia:2011tj,Nason:2013ydw} production were simulated at next-to-leading order~(NLO) using {\tt POWHEG~BOX}~\cite{Alioli:2010xd}. Only the final states with at least one neutrino are considered. The samples produced with {\tt POWHEG~BOX} are normalised to the NLO cross section given by the generator, except for $t\bar{t}$ production which is normalised to the  cross section obtained at next-to-next-to-leading order plus next-to-next-to-leading logarithmic accuracy~\cite{Czakon:2011xx,Czakon:2013goa}.

The most important experimental objects in our analysis are hadronic jets and $E_{T, \rm miss}$,  whereas charged leptons are only used for vetoing purposes. Charged leptons are constructed from stable particles in the generator output, while jets are built by clustering the true momenta of all particles but muons that interact in the calorimeters. {\tt FastJet}~\cite{Cacciari:2011ma} is used to construct anti-$k_t$ jets~\cite{Cacciari:2008gp} of radius $R=0.4$.  The variable $\vec p_{T,\rm miss}$ with magnitude $E_{T, \rm miss}$ is defined at truth level,~i.e.~before applying detector effects, as the vector sum of the transverse momenta of all  invisible particles. The effect of the detector is simulated by applying Gaussian smearing functions to the momenta of the experimental objects and by employing reconstruction and tagging efficiency factors tuned to reproduce the performance of the ATLAS detector~\cite{Aad:2008zzm,Aad:2009wy}. To smear $\etmiss$ the transverse momenta of xunsmeared electrons, muons and jets are subtracted from the truth \etmiss and replaced by the corresponding smeared quantities.  The residual truth imbalance  is then smeared as a function of the scalar sum of the $p_T$ of the particles not assigned to electrons or jets.  Similar techniques for fast detector simulation are used for the projection studies~\cite{ATL-PHYS-PUB-2016-026}  of the ATLAS collaboration, and  have also been employed in the phenomenological analyses~\cite{Haisch:2016gry,Haisch:2018tw,Pani:2017qyd,Haisch:2018djm}.

\section{Analysis strategy}
\label{sec:analysis}

\begin{table}[t!]
\begin{center}
\begin{tabular}{|l|c|c|}
\hline 
&& \\[-4.5mm]
& ${\rm SR}_{j}$ & ${\rm SR}_{2j}$ \\[0.5mm]
\hline
& \multicolumn{2}{|c|}{}  \\[-4mm] 
$\etmiss$ & \multicolumn{2}{|c|}{$> 350 \, {\rm GeV}$}\\
$\Delta\phi_{{\ptmiss} j}$ & \multicolumn{2}{|c|}{$> 0.4$} \\ 
& \multicolumn{2}{|c|}{} \\[-4mm]
\hline
& &  \\[-4mm]
leading jet & $|\eta_{j_1}|<2.4$\,, \;\; $p_{T,j_1}> 250 \, {\rm GeV}$ &  $|\eta_{j_1}|<2.4 \,, \;\; p_{T,j_1}>100 \, {\rm GeV}$ \\
& &  \\[-2mm]
subleading jet  & $\begin{cases} \; \text{n/a} \,, & N_j = 1 \,, \\ \; |\eta_{j_2}|<2.8 \,, \;\; p_{T,j_2}>30 \, {\rm GeV}   \,, & N_j > 1 \end{cases}$  &  $|\eta_{j_2}|<2.8 \,, \;\; p_{T,j_2}>50 \, {\rm GeV}$   \\
& &  \\[-2mm]
$m_{j_1 j_2} $ & $\begin{cases} \; \text{n/a} \,, & N_j = 1 \,, \\ \;  < 500 \, {\rm GeV} \, (800 \, {\rm GeV}) \,, & N_j > 1 \end{cases}$ & $> 500 \, {\rm GeV} \, (800 \, {\rm GeV})$  \\[-4mm]
& &  \\
\hline
\end{tabular}
\vspace{4mm}
\caption{Definition of the two signal regions ${\rm SR}_{j}$ and ${\rm SR}_{2j}$. See the text for additional explanations.}
\label{tab:srdef}
\end{center}
\end{table}

In our analysis, two orthogonal signal regions are defined. One focusing on the signature with a single jet~(${\rm SR}_j$), and another one with two jets of high dijet invariant mass $m_{j_1 j_2}$~(${\rm SR}_{2j}$). The definitions of the  signal regions are summarised in Table~\ref{tab:srdef}. The basic selections for both signal regions require $\etmiss>350 \, {\rm  GeV}$, and that the separation in the azimuthal angle $\Delta\phi_{ {\ptmiss} j}$  between~${\ptmiss}$ and any jet satisfies $\Delta\phi_{ {\ptmiss} j} >0.4$.   Reconstructed jets have to have  $|\eta_j| < 2.8$ and $p_{T,j} > 20 \, {\rm GeV}$, and events containing more than four jets,~i.e.~$N_j > 4$, with $p_{T,j} > 30 \, {\rm GeV}$ are vetoed. The latter two cuts ensure that the background from QCD multijet production is subdominant in the experimental analysis. We also veto events with electrons or muons.

Jets are treated differently in the two signal regions. In  ${\rm SR}_j$ (${\rm SR}_{2j}$), we demand the presence of at least one  jet ($j_1$) with  $|\eta_{j_1}|<2.4$ and $p_{T,j_1}>250 \, {\rm GeV}$ ($p_{T,j_1}> 100 \, {\rm GeV}$). Notice that the~$\etmiss$, the $\Delta\phi_{{\ptmiss} j}$ and the leading-jet cuts imposed in  ${\rm SR}_j$ match those of the  signal region~IM3 defined in~\cite{Aaboud:2017phn}.  If ${\rm SR}_j$ contains two or more jets we require $|\eta_{j_2}|<2.8$ and $p_{T,j_2}>30 \, {\rm GeV}$, whereas for~${\rm SR}_{2j}$ the  second hardest jet  has to satisfy  $|\eta_{j_2}|<2.8$ and $p_{T,j_2}>50 \, {\rm GeV}$. In the case of the signal region ${\rm SR}_{2j}$, we finally ask that the invariant mass of the two leading jets fulfills $m_{j_1j_2}> 500 \, {\rm GeV}$ ($m_{j_1j_2}> 800 \, {\rm GeV}$) in the extrapolation to $300 \, {\rm fb}^{-1}$ ($3 \, {\rm ab}^{-1}$) of integrated luminosity, while if an event in ${\rm SR}_{j}$ features more than one jet, $m_{j_1j_2}$ is required to be smaller than the cut employed in ${\rm SR}_{2j}$. We emphasise that the~${\rm SR}_{2j}$ requirements on~$m_{j_1j_2}$ have been optimised in our analysis to provide the best possible separation between  DM signal and SM background in terms of $|\Delta\phi_{j_1j_2}|$ distributions. Such an optimisation has not been performed in the earlier study~\cite{Haisch:2013fla}. 

After applying the above cuts, the SM background amounts to approximately 595\hspace{0.25mm}k~(102\hspace{0.25mm}k) events in ${\rm SR}_j$~(${\rm SR}_{2j}$) per~$300 \, {\rm fb}^{-1}$~of integrated luminosity at the 14~TeV LHC. In both signal regions,  the ratio of the number of DM signal to  SM background events turns out to be in the ballpark of~1\%~(2\%)  for a scalar (pseudoscalar) mediator with mass below $400 \, {\rm GeV}$. For~larger   values of $M_{\phi/a}$ the signal-to-background ratio  rapidly  decreases.

\begin{figure}[t!]
\begin{center}
\includegraphics[width=0.95\textwidth]{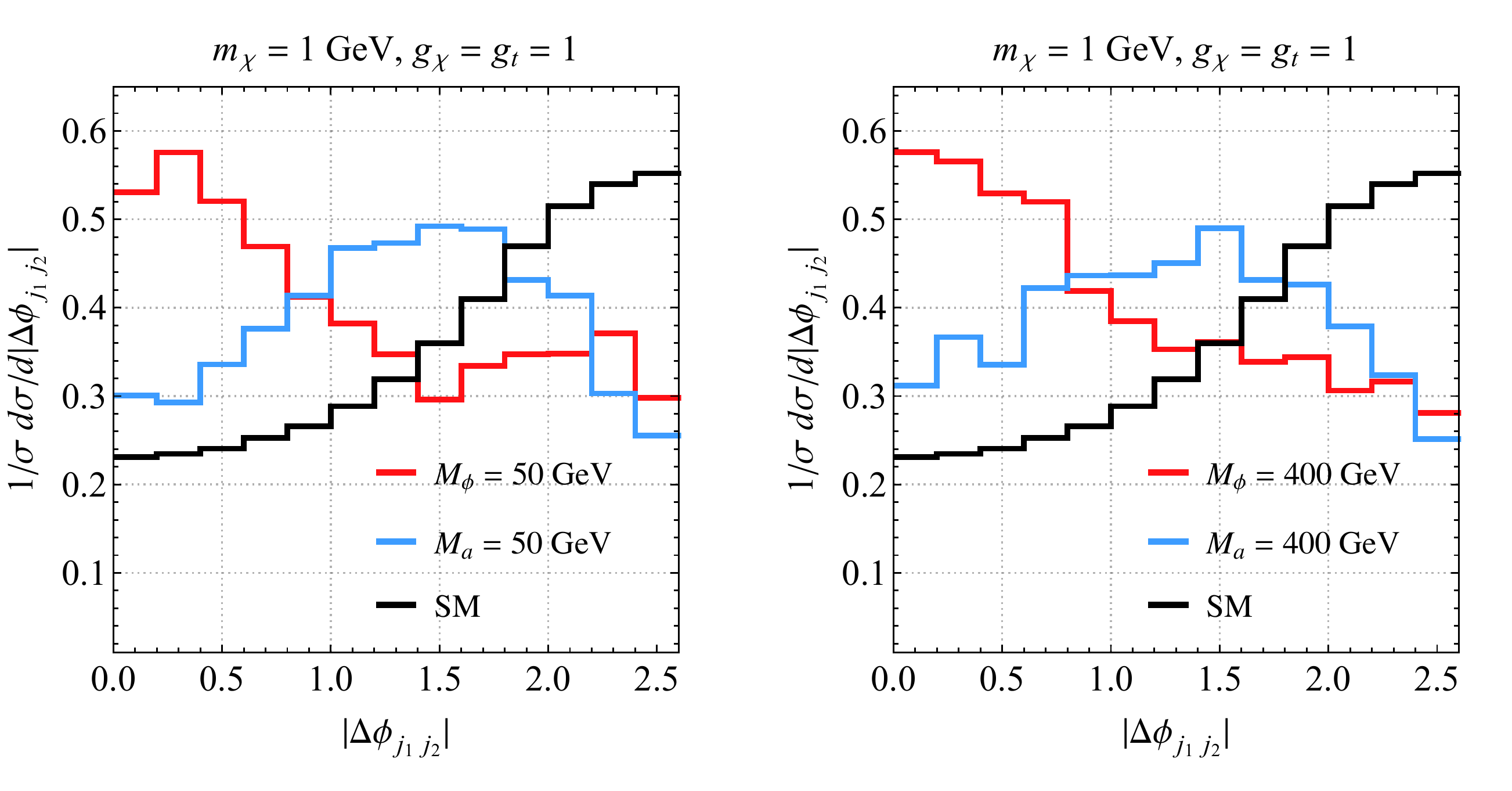}
\vspace{-2mm}
\caption{Normalised $|\Delta\phi_{j_1j_2}|$ distributions of the $2 j + \etmiss$ signature and the total SM background at the 14~TeV LHC. The shown results impose the experimental selections ${\rm SR}_{2j}$ with $m_{j_1 j_2} > 500 \, {\rm GeV}$ (cf.~Table~\ref{tab:srdef}). The choices of  the spin-0 $s$-channel DM simplified model parameters are indicated in the legends and the headlines of the two panels.  For further explanations see the main text.}
\label{dphij1j2}
\end{center}
\end{figure}
 
The distributions for the azimuthal angle difference between the two jets $|\Delta\phi_{j_1j_2}|$ in  the signal region ${\rm SR}_{2j}$ with $m_{j_1 j_2} > 500 \, {\rm GeV}$ are shown in  Figure~\ref{dphij1j2}. All the distributions are normalised to unity when integrated over $|\Delta\phi_{j_1j_2}| \in [0,2.6]$. We see that in the case of the DM signal the normalised $|\Delta\phi_{j_1j_2}|$ spectra display a pronounced cosine-like and sine-like modulation that, as explained~for example~in~\cite{Haisch:2013fla,Plehn:2001nj,Klamke:2007cu}, is typical for scalar~(red) and pseudoscalar~(blue) exchange. Compared to the $2 j +\etmiss$ scalar and pseudoscalar distributions the SM background (black) is peaked towards $|\Delta\phi_{j_1j_2}| = \pi$. The observed shape differences  offer the possibility of improving the sensitivity of the mono-jet analysis and, in the case of discovery, may allow to distinguishing  between a CP-even and a CP-odd mediation mechanism. It is important to notice in this respect that the shape of the $|\Delta\phi_{j_1j_2}|$ spectra is rather insensitive to the mediator mass as illustrated by the left and right panel corresponding to masses of $50 \, {\rm GeV}$ and $400 \, {\rm GeV}$, respectively.  We stress that compared to the previous work~\cite{Haisch:2013fla} that used hard matrix elements only, the Monte Carlo~(MC) modelling  of the DM signals performed here is more sophisticated  as it  includes  the effects of CKKW jet matching, parton shower and hadronisation corrections and a realistic detector simulation~(cf.~Section~\ref{sec:generation}). Our study thus shows that the azimuthal angle difference $\Delta\phi_{j_1j_2}$  in $2 j + \etmiss$ production is not washed out by soft physics and/or detector effects, and therefore  provides  a powerful model discriminant in realistic  LHC mono-jet analyses. In Appendix~\ref{sec:appendix} we quantify the gain in sensitivity that is achieved by adding shape information to the search strategy~${\rm SR}_{2j}$. 

Before discussing our LHC Run-3 and HL-LHC projections, we briefly comment on the relevance of the  $t \bar t$, $tW$ and diboson backgrounds in our analysis. In the signal region ${\rm SR}_{j}$, we find that the impact of  the sum of $t \bar t$ and $tW$ (diboson) production is negligible as this contribution amounts to a fraction of only 0.2\% (0.8\%) of the total SM background. In ${\rm SR}_{2j}$ the contribution due to $t \bar t$ and $tW$  (diboson) production  instead amounts to 6\% (3\%).  The  $|\Delta \phi_{j_1 j_2}|$ distribution of the $t \bar t$ and $tW$ background is however peaked at around $|\Delta \phi_{j_1 j_2}| = 2.8$ with a flat tail that slowly grows from $|\Delta \phi_{j_1 j_2}| =0$, whereas the  shape of the $|\Delta \phi_{j_1 j_2}|$ spectrum of the diboson background resembles that of the leading SM background from $Z/W + {\rm jets}$ production (cf. the black histograms in Figure~\ref{dphij1j2}). The discriminating power of the $\Delta \phi_{j_1 j_2}$ observable is therefore not affected by the subleading backgrounds if the $\Delta\phi_{j_1j_2}$ shape fit is limited to the range $|\Delta\phi_{j_1j_2}| \in [0,2.6]$, as done in the subsequent numerical analysis. 

\section{LHC Run-3 and HL-LHC projections}
\label{sec:results} 

The goal of this section is to derive upper limits on the  signal strength $\mu$,~i.e.~the ratio of the signal yield to that predicted in the spin-0 $s$-channel  DM simplified models~(\ref{eq:lagrangians}). Given that in both signal regions the signal-to-background ratio is at the  percent level, a  shape fit to a discriminant variable is necessary  to maximise the sensitivity to the DM signal. In our analysis, we perform a standard~$\etmiss$ shape fit   in  the signal region ${\rm SR}_j$, while in ${\rm SR}_{2j}$ the shape of the $|\Delta\phi_{j_1j_2}|$ distributions is used as a discriminator. In the case of the HL-LHC, the high number of events in the two signal regions implies that the sensitivity of the search largely depends on  how well the SM background can be modelled and/or constrained. Since the  systematic uncertainties plaguing the $Z/W+{\rm jets}$  background have been identified as the  limiting factor in mono-jet analyses,  much experimental and theoretical effort went into minimising these uncertainties by employing techniques that involve a mix of data-driven methods and MC studies~(see~e.g.~\cite{Aaboud:2017phn,Sirunyan:2017jix,Lindert:2017olm}).  Since it is beyond the scope of this work to perform such an evaluation on our MC generated SM backgrounds, we will rely on published experimental data to approximate the effect of systematic uncertainties on the LHC sensitivity prospectives. 

Given that apart from the additional $m_{j_1 j_2}$ cut in the $N_j > 1$ case, our signal region ${\rm SR}_j$ resembles the requirements of the  selection IM3 of the ATLAS search~\cite{Aaboud:2017phn},  we  base our extrapolations on the information provided in the latter  article. In the recent mono-jet analysis of the~ATLAS collaboration, the systematic uncertainties are evaluated through a combined shape fit to the signal  region and to appropriate control regions enriched by the dominant  SM backgrounds. The obtained systematic uncertainty on the number of  SM background events in  IM3 amounts to~2.6\%, and we assume that  uncertainties of the same size also arise in the case of our signal regions ${\rm SR}_j$ and ${\rm SR}_{2j}$. Besides the systematic uncertainty on the normalisation of the SM background, also the shapes of the distributions which enter the likelihood fits carry uncertainties. In the case of~\cite{Aaboud:2017phn} for instance,  the bin-by-bin systematic uncertainties on the shape of the $\etmiss$ distribution amount to around~$[3\%, 7\%]$ for  $\etmiss \in [350 \, {\rm GeV}, 1 \, {\rm TeV}]$. These uncertainties are however strongly correlated among bins and cannot be naively  used as bin-by-bin errors in a likelihood fit. In~addition, no experimental information on the  systematic uncertainties of the $\Delta\phi_{j_1j_2}$ distributions  is available, although it seems likely  that the shapes of the spectra shown in Figure~\ref{dphij1j2} can be modelled with higher  precision than the steeply falling \etmiss distributions. As we are mainly interested in the relative reach of the two signal regions ${\rm SR}_j$ and ${\rm SR}_{2j}$, as a minimal approach we only use the uncertainty of~$2.6\%$ on the total number of expected events in each signal region, ascribing no additional error to the shapes of the $\etmiss$ and $\Delta\phi_{j_1j_2}$ distributions. This procedure will  allow us to calculate an upper limit on the sensitivity of our analysis strategy. 

\begin{figure}[t!]
\begin{center}
\includegraphics[width=0.95\textwidth]{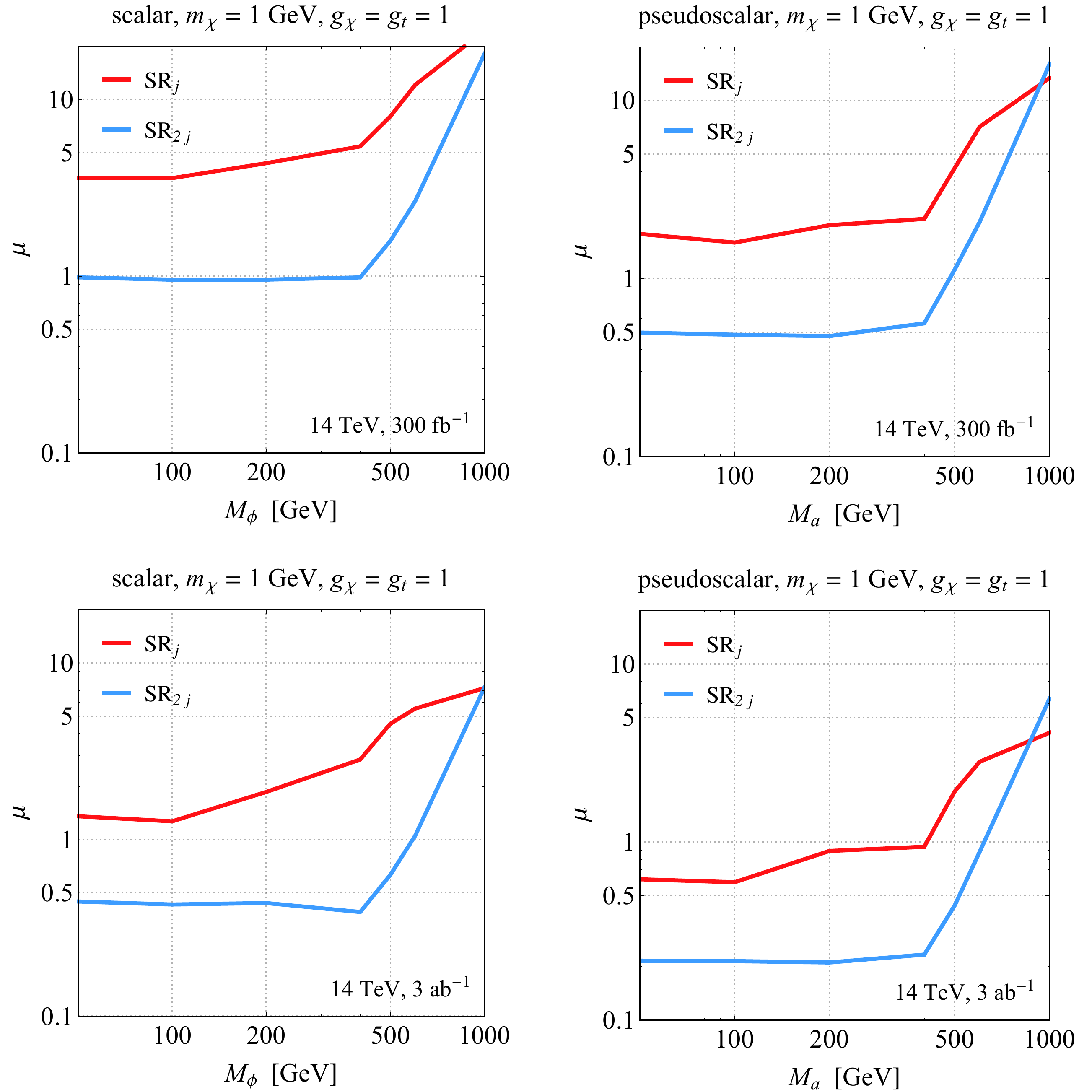}
\vspace{2mm}
\caption{Value of the signal strength $\mu$ that can be excluded at 95\% CL as a function of the mass for scalar (left) and pseudoscalar (right) mediators. The reach for $300 \, {\rm fb}^{-1}$ (upper row) and $3 \, {\rm ab}^{-1}$ (lower row) of 14~TeV LHC data is shown for the parameter choices $m_\chi = 1 \, {\rm GeV}$ and $g_\chi = g_t = 1$. The displayed results assume an uncertainty of~$2.6\%$ on the total number of expected events in both signal regions.  Consult the text for additional explanations.}
\label{reach26dmj}
\end{center}
\end{figure}

In order to evaluate the upper confidence level (CL) limits on the signal strength $\mu$, we construct parametrised probability density functions for the DM signals and the SM background with  the {\tt HistFactory} package~\cite{Cranmer:2012sba}. The significance is then calculated using  the $C\hspace{-0.2mm}L_s$ method~\cite{Read:2002hq}. The actual calculation is performed with the {\tt RooStats} toolkit~\cite{Moneta:2010pm}, which utilises the asymptotic formulas for likelihood-based tests presented in \cite{Cowan:2010js}. The assumptions on systematic uncertainties incorporated in the probability density functions for each of the signal regions has already been discussed before. Figure~\ref{reach26dmj} displays our 95\% CL limits for integrated luminosities of $300 \, {\rm fb}^{-1}$~(upper row) and~$3 \, {\rm ab}^{-1}$~(lower row) as a function of the mediator mass for scalar~(left) and pseudoscalar~(right) mediators. The red (blue) curves correspond to the results of the \etmiss ~($\Delta\phi_{j_1j_2}$) shape fit in ${\rm SR}_j$ (${\rm SR}_{2j}$) as described above.   The corresponding fit ranges are $\etmiss > 350 \, {\rm GeV}$ and $|\Delta\phi_{j_1j_2}| \in [0,2.6]$, respectively.   One observes that under the assumption of systematic uncertainties of $2.6\%$ on the number of  SM background events in  ${\rm SR}_j$ and ${\rm SR}_{2j}$, the $\Delta\phi_{j_1j_2}$ shape fit proposed by us leads to significantly stronger LHC Run-3 and the HL-LHC constraints on $\mu$ than a standard \etmiss shape analysis. This finding can be understood qualitatively by recalling that there is a large fraction of two jet-like events in the case of gluon-fusion induced mono-jet production~\cite{Haisch:2013ata}, and our $\Delta\phi_{j_1j_2}$ shape fit exploits this feature. Numerically, our ${\rm SR}_{2j}$ search strategy leads to the 95\%~CL  limits $M_{\phi} > 402 \, {\rm GeV}$ and $M_{a} > 477\, {\rm GeV}$ for $m_\chi = 1 \, {\rm GeV}$, $g_\chi = g_t = 1$ and $300 \, {\rm fb}^{-1}$ of 14~TeV data.  The corresponding bounds for $3 \, {\rm ab}^{-1}$ of integrated luminosity read $M_{\phi} > 587 \, {\rm GeV}$ and $M_{a} > 608\, {\rm GeV}$.  We emphasise that the quoted exclusions have been obtained under the assumption that only the  total number of expected events carries a systematic uncertainty. This procedure hence  leads to  upper bounds on the sensitivity of our analysis strategy~${\rm SR}_{2j}$, corresponding to the limit of $\Delta\phi_{j_1j_2}$ distributions with vanishing  shape uncertainties. Since  a reliable estimate of shape uncertainties and their correlations is only possible in an analysis that uses real LHC data, it is beyond the scope of this work to quantify to which extent our projections would be weakened if shape uncertainties were to be included. Additional extrapolations based on a different systematic uncertainty scenario can be found in Appendix~\ref{sec:appendix}.

\section{Conclusions and outlook}
\label{sec:conclusions}

The main goal of this article was  to put the earlier study \cite{Haisch:2013fla} of angular correlations in gluon-fusion production of loop-induced $2j + \etmiss$ signatures on more solid ground both from a theoretical and experimental point of view. To this purpose, we have performed state-of-the-art MC simulations of both the mono-jet signal in spin-0 $s$-channel simplified models and the associated SM backgrounds. The dominant background from $Z/W +{\rm jets}$ production, but also the subleading $t \bar t$, $tW$ and diboson channels have been considered. Our event generation includes  the effects of jet matching and merging as well as parton shower and hadronisation corrections, and we have performed a realistic detector modelling (see Section~\ref{sec:generation}). 

The proposed  analysis strategy aims at separating the mono-jet signature into two distinct signal regions. The first signal region called ${\rm SR}_j$ focuses on single jet-like events, while the second signal region referred to as ${\rm SR}_{2j}$ requires the presence of a  jet pair with large invariant mass in the final state~(see~Section~\ref{sec:analysis}). In the signal region ${\rm SR}_{2j}$, we have studied the azimuthal angle difference $\Delta\phi_{j_1j_2}$  in mono-jet production. Our study shows  that the latter observable provides a powerful model discriminant in realistic  LHC analyses. In fact,  shape fits to the $\Delta\phi_{j_1j_2}$ variable will generically help to  improve the sensitivity of mono-jet searches (see Appendix~\ref{sec:appendix}), and, in the case of a discovery, might allow to distinguishing between scalar and pseudoscalar mediators. 

We have then analysed the mono-jet coverage of the parameter space of   the spin-0 $s$-channel  DM simplified models expected at LHC~Run-3 and the HL-LHC~(see~Section~\ref{sec:results}). In a first step, we have derived hypothetical limits on the signal strength that follow from a standard $\etmiss$ shape analysis to the ${\rm SR}_j$ signal region. In a second step, we have then obtained bounds by performing shape fits to the $\Delta\phi_{j_1j_2}$ variable utilising the ${\rm SR}_{2j}$ event samples. Under the reasonable assumption that the systematic uncertainties on the number of  SM background events in  ${\rm SR}_j$ and ${\rm SR}_{2j}$ are the same, we have found that the proposed $\Delta\phi_{j_1j_2}$ shape fit has a significantly better reach  than a standard \etmiss shape analysis.  For the  benchmark parameter choices $m_\chi = 1 \, {\rm GeV}$ and $g_\chi = g_t = 1$, the 95\%~CL exclusion limits that derive from our ${\rm SR}_{2j}$ search strategy read $M_{\phi} > 402 \, {\rm GeV}$ and $M_{a} > 477\, {\rm GeV}$ for $300 \, {\rm fb}^{-1}$ of 14~TeV data.  The corresponding bounds for $3 \, {\rm ab}^{-1}$ of integrated luminosity turn out to be $M_{\phi} > 587 \, {\rm GeV}$ and $M_{a} > 608\, {\rm GeV}$. Notice that in the scalar case the quoted LHC Run-3 limit  is slightly weaker than the bound that follows from a  combined analysis of $t\bar t +\etmiss$ and $t W +\etmiss$ production~\cite{Haisch:2018tw}, while in all other cases the mono-jet  sensitivity exceeds that of the $tX+\etmiss$ search.  This finding  illustrated the synergy and complementarity of the latter two mono-$X$ channels~\cite{Buckley:2014fba,Haisch:2015ioa} in the context of spin-0 $s$-channel DM simplified models. 

We finally note that the ${\rm SR}_{2j}$ analysis strategy proposed by us can also be straightforwardly  applied to next-generation DM simplified models such two-Higgs-doublet extensions with an extra spin-0 gauge singlet~\cite{Ipek:2014gua,No:2015xqa,Goncalves:2016iyg,Bell:2016ekl,Bauer:2017ota,Tunney:2017yfp,Abe:2018bpo}. Like in the case of the spin-0 $s$-channel DM simplified models discussed here, we expect that exploiting the $\Delta \phi_{j_1 j_2}$  correlations in $2 j + \etmiss$ production will also allow to significantly strengthen future LHC mono-jet  constraints on spin-0 next-generation DM simplified models. 

\acknowledgments 
We are grateful to  Giuliano Gustavino for    useful comments on the manuscript.

\begin{appendix}

\begin{figure}[t!]
\begin{center}
\includegraphics[width=0.95\textwidth]{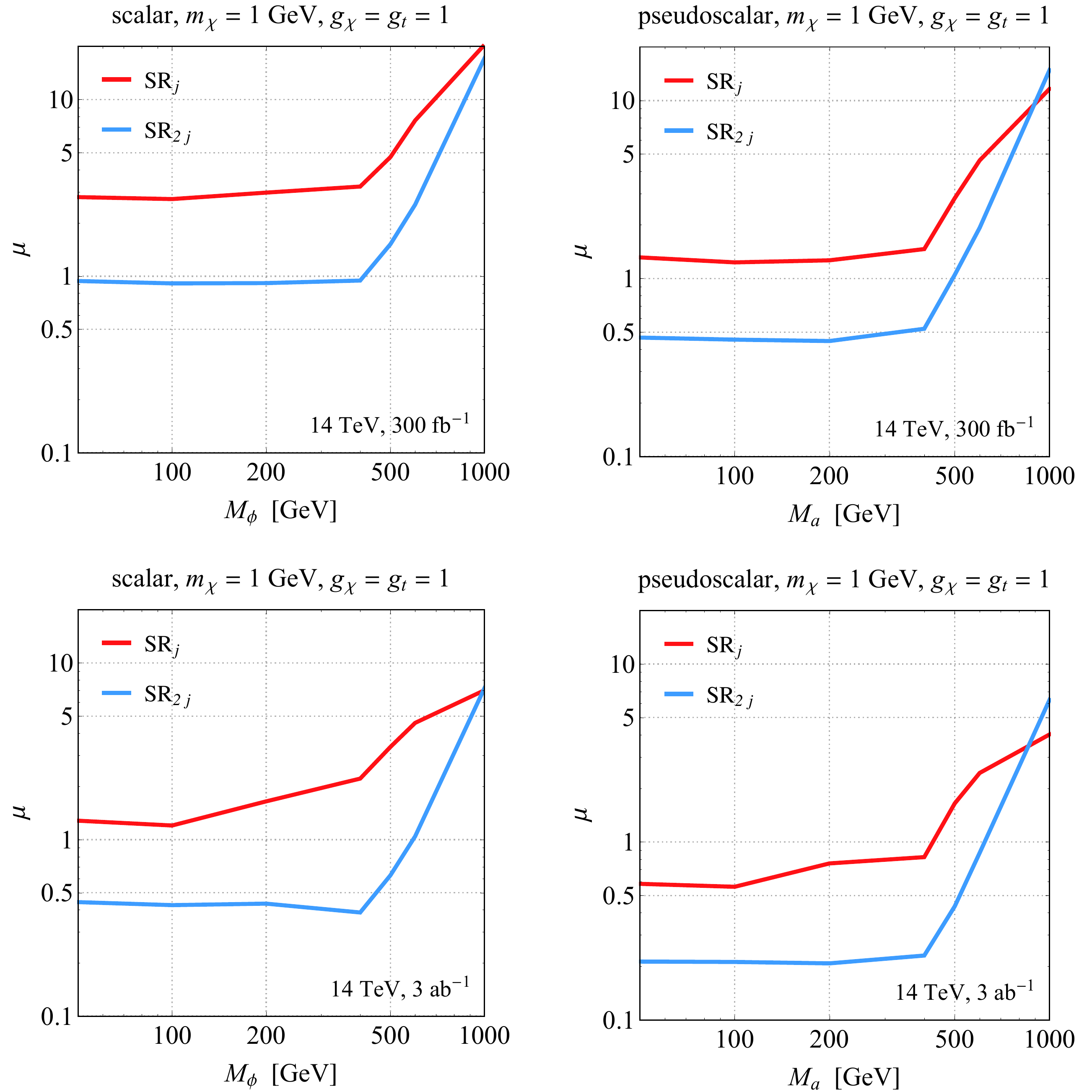}
\vspace{2mm}
\caption{As Figure~\ref{reach26dmj} but  assuming uncertainties of~$1.3\%$ on the total number of expected events in the signal regions ${\rm SR}_j$ and ${\rm SR}_{2j}$. See the  text for further explanations. }
\label{reach13dmj}
\end{center}
\end{figure}

\section{Supplementary material}
\label{sec:appendix}

In this appendix we extend the numerical study performed in Section~\ref{sec:results}. We start by presenting LHC explorations based on an alternative more aggressive assumption about the systematic uncertainties of future LHC mono-jet searches. Anticipating improvements in detector performance and modelling of DM signal and SM~background processes,   we assume, in the spirit of~\cite{CMS-PAS-FTR-16-005,ATL-PHYS-PUB-2018-043},  that the present systematic uncertainties on the   total number of expected events in the signal regions ${\rm SR}_j$ and ${\rm SR}_{2j}$ can be reduced by a factor of~2. In Figure~\ref{reach13dmj} we show the 95\% CL limits for $300 \, {\rm fb}^{-1}$~(upper row) and~$3 \, {\rm ab}^{-1}$~(lower row) of data as a function of  the scalar~(left) and pseudoscalar~(right) mediator mass.   The red (blue) curves illustrate the results of the \etmiss ~($\Delta\phi_{j_1j_2}$) shape fit in ${\rm SR}_j$~(${\rm SR}_{2j}$) as described in Section~\ref{sec:results}, assuming an improved systematic uncertainty of $1.3\%$. Under this assumption, we find that the proposed ${\rm SR}_{2j}$ search strategy leads to the 95\%~CL  limits $M_{\phi} > 409 \, {\rm GeV}$ and $M_{a} > 490 \, {\rm GeV}$ for $m_\chi = 1 \, {\rm GeV}$, $g_\chi = g_t = 1$ and $300 \, {\rm fb}^{-1}$ of 14~TeV data.  The corresponding  $3 \, {\rm ab}^{-1}$ bounds are $M_{\phi} > 589 \, {\rm GeV}$ and $M_{a} >  609 \, {\rm GeV}$. Notice that these limits are only marginally better than the bounds reported at the end of Section~\ref{sec:results}. The 95\%~CL bounds on~$\mu$ that derive from the search strategy ${\rm SR}_j$ are in contrast  notable  improved  if the systematic uncertainties are reduced from 2.6\% to 1.3\%. Numerically, we find average improvements of 45\% and 15\% at LHC Run-3 and HL-LHC, respectively.

\begin{figure}[t!]
\begin{center}
\includegraphics[width=0.95\textwidth]{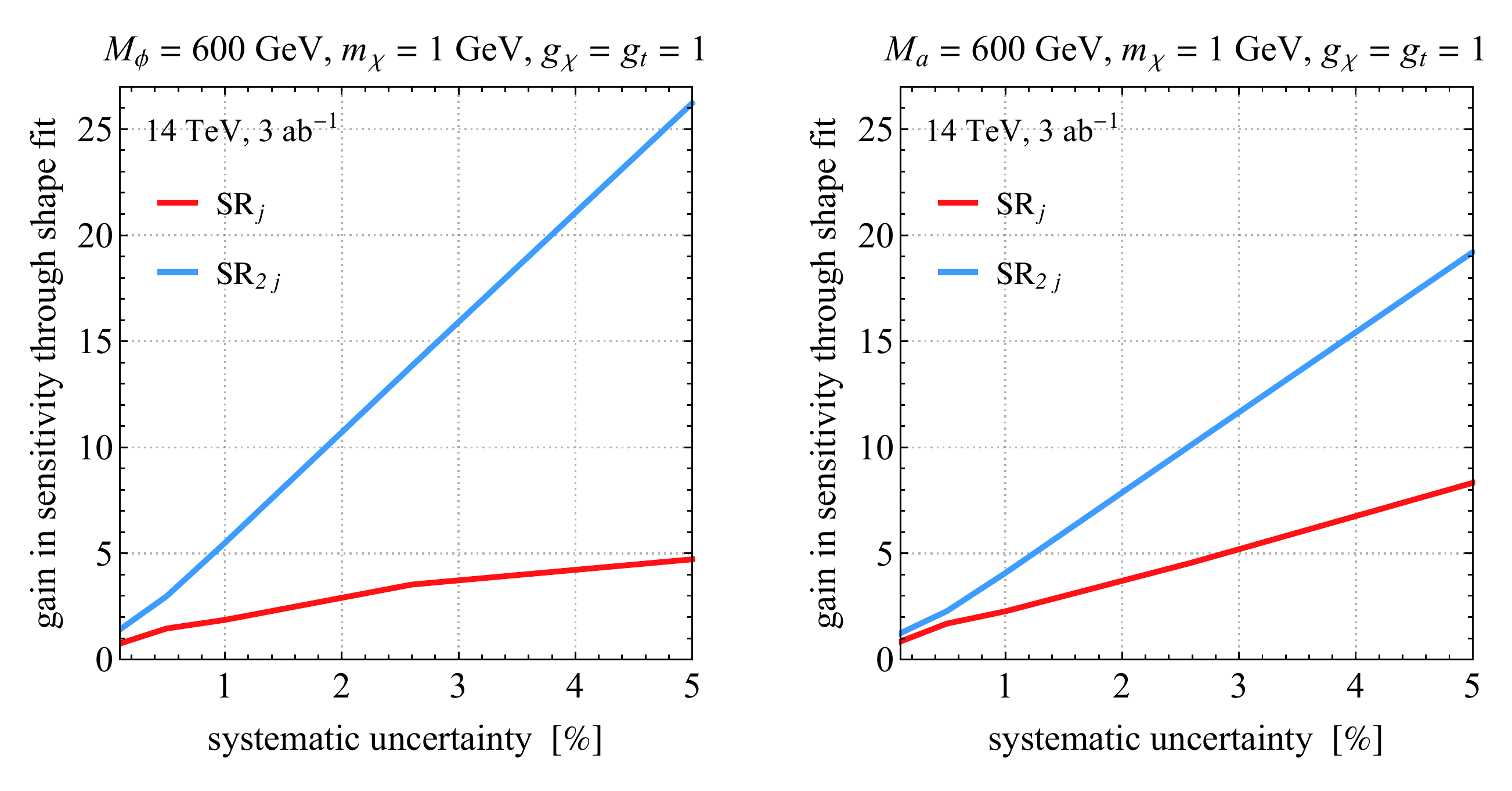}
\vspace{0mm}
\caption{Gain in sensitivity that is achieved by adding shape information to the search strategy ${\rm SR}_j$~(red~curves) and ${\rm SR}_{2j}$~(blues~curves) as a function of the assumed systematic uncertainties in percent. The shown results correspond to $3 \, {\rm ab}^{-1}$ of 14~TeV LHC data and the used parameter choices are indicated in the headlines of the two panels. See text for additional information.}
\label{shape}
\end{center}
\end{figure}

\begin{figure}[t!]
\begin{center}
\includegraphics[width=0.95\textwidth]{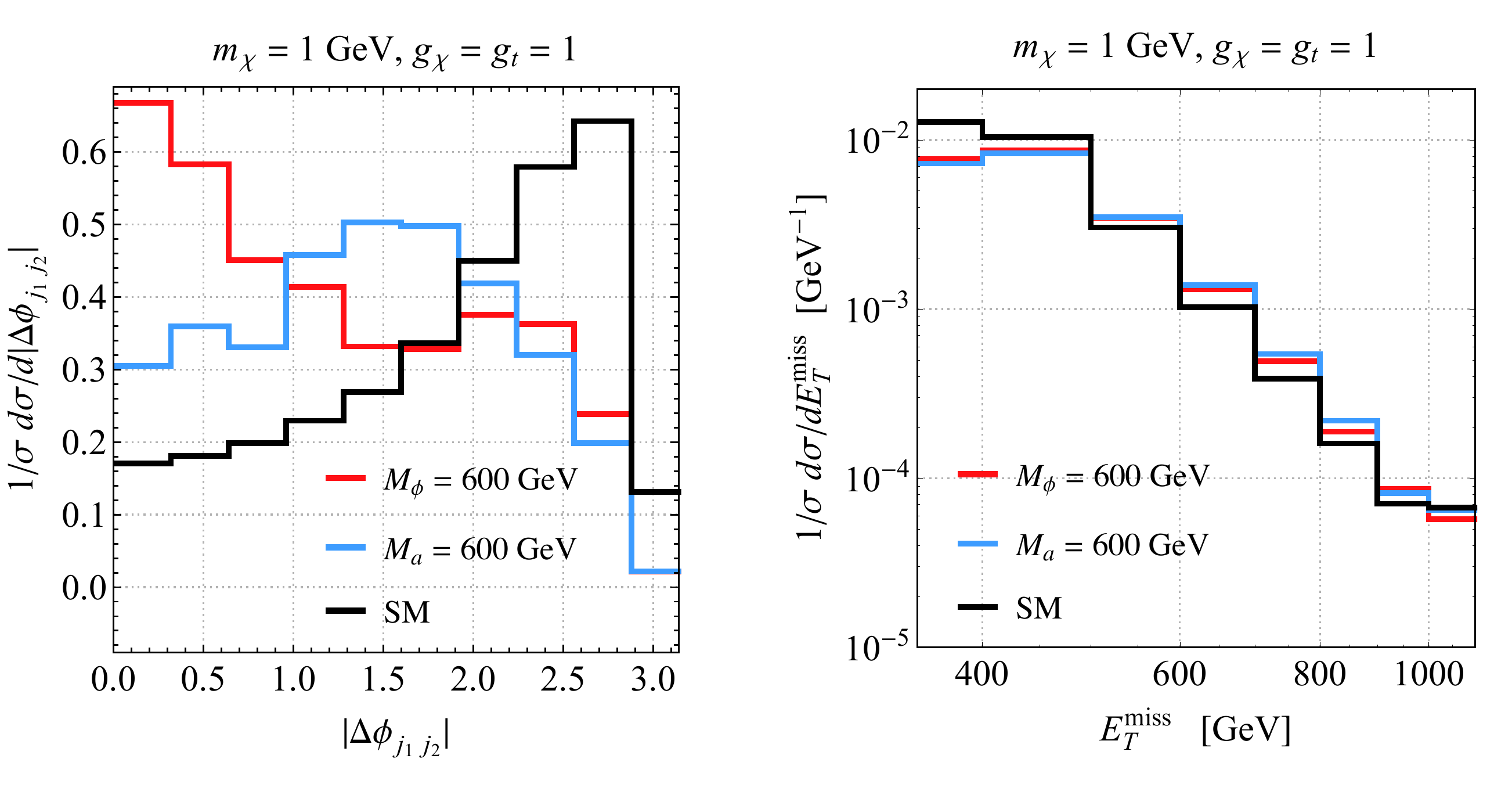} 
\vspace{0mm}
\caption{Normalised $|\Delta\phi_{j_1j_2}|$ (left) and $\etmiss$ (right) distributions in the  ${\rm SR}_{2j}$ and  ${\rm SR}_{j}$, respectively. The last bin of the $\etmiss$ histograms is an overflow bin. The shown results correspond to $3 \, {\rm ab}^{-1}$ of 14~TeV~data  and they impose the HL-LHC cuts as specified in Table~\ref{tab:srdef}. The~red~(blue) histograms represent the DM signal arising from scalar (pseudoscalar) exchange, while the SM background distributions are coloured black.  The legends and the headlines of the panels  indicate the used spin-0 $s$-channel DM simplified model parameters.}
\label{phiMETshape}
\end{center}
\end{figure}

In addition let us quantify the impact of shape information in the two mono-jet search strategies considered by us. To do so, we define the gain of sensitivity through the shape fit as the ratio of $\mu$ values obtained with and without the inclusion of shape information. This ratio is displayed  in Figure~\ref{shape} as a function of the assumed systematic uncertainty  on the number of events in ${\rm SR}_j$~(red~curves) and ${\rm SR}_{2j}$~(blue~curves). The shown results correspond to the HL-LHC and two benchmark spin-0 $s$-channel DM simplified models. From the panels it is evident that the shape information carried by  $\Delta \phi_{j_1 j_2}$ is a significantly more powerful constraint than that of $\etmiss$. This finding is unsurprising, if one considers the shapes of the $\Delta \phi_{j_1 j_2}$ and $\etmiss$ corresponding to the parameter choices used to obtain the latter figure. As can be seen from Figure~\ref{phiMETshape}, the $\Delta \phi_{j_1 j_2}$ spectrum displays a marked cosine-like (sine-like)  modulation in the scalar (pseudoscalar) case, while the $\etmiss$ distributions are steeply falling and  largely independent of the mediator type. In the case of the $\Delta \phi_{j_1 j_2}$ distributions, one furthermore observes a clear distinction between the shapes of the DM signals and the SM background, while in the \etmiss case the differences between the three normalised spectra are significantly less prominent. 

The features of the results shown in Figures~\ref{reach26dmj}, \ref{reach13dmj}, \ref{shape} and  \ref{phiMETshape} thus strongly suggest that search strategies based on $\Delta \phi_{j_1 j_2}$ shape fits are not only more powerful than standard $\etmiss$ shape analyses in constraining the parameter space of spin-0 $s$-channel DM simplified models, but are also less dependent on hypothetical improvements  of the systematic uncertainties of future mono-jet searches. 

\end{appendix}

%\bibliographystyle{JHEP}
%\bibliography{2jMET}

\providecommand{\href}[2]{#2}\begingroup\raggedright\endgroup

\end{document}